\documentclass[a4paper,11pt]{article}
\usepackage{amssymb,amsmath}
\usepackage{mathrsfs}

\allowdisplaybreaks
\numberwithin{equation}{section}
\newtheorem{thm}{Theorem}[section]
\newtheorem{prop}[thm]{Proposition}

\newtheorem{defn}[thm]{Definition}

\newtheorem{cor}[thm]{Corollary}
\newtheorem{example}[thm]{Example}

\newtheorem{remark}[thm]{Remark}

\newcommand\tr{\mathrm{tr}}

\newcommand{\diag}{\mathrm{diag}}

\newcommand\pd{\partial}
\newcommand{\ld}{\lambda}
\newcommand{\al}{\alpha}

\newcommand{\sg}{\sigma}
\newcommand{\om}{\omega} 

\newcommand{\Ld}{\Lambda}

\newcommand{\dt}{\delta}

\newcommand{\res}{\mathrm{res}}

\newcommand\C{\mathbb{C}}

\newcommand\Z{\mathbb{Z}}

\newcommand\mfg{\mathfrak{g}}
\newcommand\mfh{\mathfrak{h}}

\newcommand\mfs{\mathfrak{s}}
\newcommand\mr{\mathring}

\newcommand\ra{\rangle}
\newcommand\la{\langle}

\begin{document}
\title{A Remark on Kac-Wakimoto Hierarchies of D-type}
\author{Chao-Zhong Wu\thanks{wucz05@mails.tsinghua.edu.cn} \\
{\small Department of Mathematical Sciences, Tsinghua University, }\\
{\small Beijing 100084, P. R. China}}
\date{}
\maketitle

\begin{abstract}
For the Kac-Wakimoto hierarchy constructed from the principal vertex
operator realization of the basic representation of the affine Lie
algebra $D_n^{(1)}$, we compute the coefficients of the
corresponding Hirota bilinear equations, and verify the coincidence
of these bilinear equations with the ones that are satisfied by
Givental's total descendant potential of the $D_n$ singularity, as
conjectured by Givental and Milanov in \cite{GM}.

\end{abstract}

\section{Introduction}
The theory on representation theoretical aspects of soliton
equations developed by  Date, Jimbo, Kashiwara,
Miwa~\cite{DJKM-KP}-\cite{DJKM-reduce} and Kac, Wakimoto~\cite{Kac,
KW} plays a significant role in several research areas of modern
mathematical physics. For each affine Lie algebra $\mfg$ together
with an integrable highest weight representation $V$ of $\mfg$ and a
vertex operator construction $R$ of $V$, Kac and Wakimoto formulated
a hierarchy of soliton equations. These equations can be written
down in terms of Hirota bilinear equations and their super analogue
\cite{KW}. When $\mfg$ is the untwisted affinization of a simply
laced finite Lie algebra, the Kac-Wakimoto hierarchy coincides with
the corresponding generalized Drinfeld-Sokolov hierarchy defined by
Groot, Hollowood and Miramontes \cite{GHM, HM}. In particular, if
the highest weight representation is the basic one, and the vertex
operator realization is constructed from the principal Heisenberg
subalgebra, then the Kac-Wakimoto hierarchy is equivalent to the
Drinfeld-Sokolov hierarchy associated to $\mfg$ and the vertex $c_0$
of its Dynkin diagram \cite{DS}.

In \cite{Given, Gi-GW}, Givental constructed the total descendant
potential for any semisimple Frobenius manifold \cite{Du}. This
potential is supposed to satisfy the axioms dictated by
Gromov-Witten theory, such as the string equation, dilaton equation,
topological recursion relations, and Virasoro constraints. Recently
Givental and Milanov \cite{Gi-A, GM} showed that the total
descendant potentials for semisimple Frobenius manifolds associated
to simple singularities satisfy certain Hirota bilinear (quadratic)
equations, and proved that for the $A_n$, $D_4$ and $E_6$
singularities these equations are equivalent to the corresponding
Kac-Wakimoto hierarchies. They also conjectured that this fact is
true for all simple singularities.

In this note we compute explicitly the coefficients of the
Kac-Wakimoto hierarchy constructed from the principal vertex
operator realization of the basic representation of the affine Lie
algebra $D_n^{(1)}$, while these coefficients are implicitly defined
in \cite{KW} except for the case $n=4$. This computation verifies
Givental and Milanov's conjecture for the $D_n$ singularity.

\section{Kac-Wakimoto hierarchies of ADE-type}
Let~$\mfg$ be an untwisted affine Lie algebra of ADE-type, with
rank~$n$, Coxeter number~$h$, and normalized invariant bilinear
form~$(\cdot\mid\cdot)$. The set of simple roots and simple coroots
are denoted by~$\{\al_i\}_{i=0}^n$ and~$\{\al_i^\vee\}_{i=0}^n$
respectively.

We denote the principal gradation of~$\mfg$ as
$\mfg=\bigoplus_{j\in\Z}\mfg_j$. The Cartan subalgebra of~$\mfg$,
i.e., the $0$-component~$\mfg_0$, has the following two
decompositions
\[
\mfg_0=\mathring{\mfh}\oplus\C c\oplus\C
d=\bar{\mfh}\oplus\C c\oplus\C d.
\]
Here on the one hand~$\mathring{\mfh}=\sum_{i=1}^n\C\al_i^\vee$, $c$
is the central element and~$d$ is determined by the constraint
\[
(\mathring{\mfh}|d)=0, ~~(c| d)=1, ~~(d| d)=0;
\]
on the other hand, the subspace~$\bar{\mfh}$ is so chosen that the
difference of the projections of any~$x\in\mfg_0$
onto~$\mathring{\mfh}$ and~$\bar{\mfh}$ is given by
$\mathring{x}-\bar{x}=h^{-1}(\mathring{\rho}^\vee|\mathring{x})c$,
where~$\mathring{\rho}^\vee$ is an element of~$\mathring{\mfh}$
defined by the condition
\begin{equation}\label{rhov}
\la\al_i,\mathring{\rho}^\vee\ra=1,\quad i=1, \dots, n.
\end{equation}

Let~$E$ be the set of exponents of~$\mfg$. For each~$j\in E$ there
exists~$H_j\in \mfg_j$ satisfying
\begin{equation}
(H_i| H_j)=h\,\dt_{i,-j},\quad [H_i, H_j]=i\,\dt_{i,-j}\,c.
\end{equation}
They generate the principal Heisenberg subalgebra~$\mathfrak{s}=\C
c+\sum_{j\in E}\C H_j$.

In Kac and Wakimoto's construction of their hierarchies, it is
essential to choose two bases~$\{v_i\}$, $\{v^i\}$ of~$\mfg$ that
are dual to each other. These two bases read
\begin{align}\label{dual}
\{v_i\}~:~& \frac{1}{\sqrt{h}}H_{j}~(j\in E), ~X^{(r)}_{m} ~(1\leq
r\leq n; m\in\Z), ~ c, ~ d; \\
\label{dual2} \{v^i\}~:~& \frac{1}{\sqrt{h}}H_{-j}~(j\in E),
~Y^{(r)}_{-m} ~(1\leq r\leq n; m\in \Z), ~ d, ~ c
\end{align}
such that
\begin{align}\label{} \label{X0Y0}
& \{X^{(r)}_0\}_{r=1}^n, \{Y^{(r)}_0\}_{r=1}^n\hbox{ are two bases
of }\bar{\mfh},\\ \label{beta}
&[H_{j},X^{(r)}_{m}]=\beta_{r,\bar{j}}X^{(r)}_{m+j},
 ~~[H_{j},Y^{(r)}_{-m}]=-\beta_{r,\bar{j}}Y^{(r)}_{-m+j},
 \\ \label{XY}
& (X^{(r)}_{l}|Y^{(s)}_{-m})=\dt_{r,s}\dt_{l,m}
\end{align}
where~$0<\bar{j}<h$ is the remainder of~$j$ modulo~$h$,
and~$\beta_{r,\bar{j}}$ are some complex numbers which depend on the choice of the two bases of $\mfg$ .

Let~$E_+$ be the set of positive exponents.
A representation of the Heisenberg subalgebra $\mr{\mfs}$ on the Fock space~$\C[t_j;\,j\in
E_+]$ is given by
\[
c\mapsto 1, ~~H_j\mapsto\frac{\pd}{\pd t_j}, ~~ H_{-j}\mapsto j\,t_j, ~~
j\in E_+.
\]
This can be lifted to a basic representation~$L(\Ld_0)$ of~$\mfg$ as
follows:
\begin{align*}\label{}
& \sum_{m\in\Z}X^{(r)}_m z^{-m}\mapsto
-h^{-1}(\mathring{\rho}^\vee|\mathring{X}^{(r)}_0)X^{(r)}(t;z), \\
& \sum_{m\in\Z}Y^{(r)}_{-m} z^{m}\mapsto
-h^{-1}(\mathring{\rho}^\vee|\mathring{Y}^{(r)}_0)X^{(r)}(-t;z),\\
&d_0:=h d+\mathring{\rho}^\vee \mapsto -\sum_{j\in E_+}j\,
t_j\frac{\pd}{\pd t_j},
\end{align*}
where $X^{(r)}(t;z)$ $(1\leq r\leq n)$ are the vertex
operators
\begin{equation*}\label{}
X^{(r)}(t;z)=\Big(\exp\sum_{j\in E_+}\beta_{r,\bar{j}}\,t_j
z^j\Big)\Big(\exp-\sum_{j\in E_+}\frac{\beta_{r,\overline{-j}}}{j
z^j}\frac{\pd}{\pd t_j}\Big).
\end{equation*}
Such a realization of the basic representation~$L(\Ld_0)$ is called
the \emph{principal vertex operator construction}, see~\cite{Kac,
KW} for details.
\begin{thm}[\cite{KW}]\label{thm-KW}
Consider the basic representation of a simply laced affine Lie
algebra~$\mfg$ on the Fock space~$L(\Ld_0)=\mathbb{C}[t_j;\,j\in
E_+]$ constructed as above. Denote by~${G}$ the Lie group of the
derived algebra $\mfg'$ of $\mfg$. A nonzero~$\tau\in L(\Ld_0)$ lies
in the orbit~$G\cdot 1$ if and only if~$\tau$ satisfies the
following hierarchy of Hirota bilinear equations:
\begin{equation}\label{kwhr}
\begin{split}
&\Big(-2h\sum_{j\in E_+}j\,y_j
D_j+\sum_{r=1}^{n}g_r\sum_{m\geq1}S_m^{E}(2\beta_{r,\bar{j}}\,y_j)S_m^{E}(-\frac{\beta_{r,\overline{-j}}}{j}D_j)\Big)\times\\
&~~\times\Big(\exp\sum_{j\in E_+}y_j D_j\Big)\tau\cdot\tau=0.
\end{split}
\end{equation}
Here~$g_r=(\mathring{\rho}^\vee|\mathring{X}^{(r)}_0)(\mathring{\rho}^\vee|\mathring{Y}^{(r)}_0)$,
$S_m^E$ are the elementary Schur polynomials of~$\mfg$ defined
by~$\exp\sum_{j\in E_+}y_j z^j=\sum_{m\geq0}S_m^{E}(y_j)z^m$,
and~$D_j$ are the Hirota bilinear operators defined by~$D_j\,
f\cdot g=\left.\frac{\pd}{\pd u}\right|_{u=0}f(t_j+u)g(t_j-u)$.
\end{thm}

Kac and Wakimoto gave explicitly the coefficients~$g_r, \beta_{r,j}$
for the affine Lie algebras $A_n^{(1)}$, $D_4^{(1)}$ and~$E_6^{(1)}$
in \cite{KW}, however, these coefficients remain implicit for other
affine Lie algebras. We proceed to compute them for the affine Lie
algebra $D^{(1)}_n$  in the next section.

\section{Bilinear equations for $D_n^{(1)}$}

Let $\mfg$ be an affine Lie algebra of type~$D_n^{(1)}$. In this
section we want to construct the two bases~\eqref{dual},
\eqref{dual2} of~$\mfg$, and then write down the Kac-Wakimoto
bilinear equations~\eqref{kwhr}. Our result implies that Givental
and Milanov's conjecture on the total descendant potential of $D_n$
singularity is true.

Let us consider the corresponding simple Lie algebra first. The
simple Lie algebra $\mr\mfg$ of type $D_n$ possesses the following
$2n$-dimensional matrix realization~\cite{DS}:
\begin{equation}\label{g}
\mathring{\mfg}=\left\{A\in\C^{2n\times 2n}\mid A=-S A^T S\right\},\ S=\sum_{i=1}^{n}(-1)^{i-1}(e_{ii}+e_{2n+1-i,2n+1-i}).
\end{equation}
Here $e_{i,j}$ is the~$2n\times 2n$ matrix that takes value $1$ at
the~$(i,j)$-entry and zero elsewhere, and $A^T=(a_{l+1-j,k+1-i})$
for any~$k\times l$ matrix~$A=(a_{ij})$. In this matrix realization,
a set of Weyl generators can be chosen as
\begin{align}\label{efhh}
&e_i=e_{i+1,i}+e_{2n+1-i,2n-i}~ (1\leq i\leq n-1), ~e_n=\frac{1}{2}(e_{n+1,n-1}+e_{n+2,n}), \\
&f_i=e_{i,i+1}+e_{2n-i,2n+1-i}~ (1\leq i\leq n-1), ~f_n={2}(e_{n-1,n+1}+e_{n,n+2}), \\
&h_i=-e_{i,i}+e_{i+1,i+1}-e_{2n-i,2n-i}+e_{2n+1-i,2n+1-i}~ (1\leq i\leq n-1), \\
&h_n=-e_{n-1,n-1}-e_{n,n}+e_{n+1,n+1}+e_{n+2,n+2}. \label{efht}
\end{align}
Besides them we also need the following elements in~$\mr{\mfg}$:
\begin{align}
&e_0=\frac{1}{2}(e_{1,2n-1}+e_{2, 2n}), ~~f_0=2(e_{2n-1,1}+e_{2n,
2}),\\
&h_0=e_{1,1}+e_{2,2}-e_{2n-1,2n-1}-e_{2n,2n}.
\end{align}
Recall the normalized Killing form $(A|B)=\frac{1}{2}\tr\,(AB)$ and
the Coxeter number~$h=2n-2$  of~$\mr\mfg$. We denote the
$\Z/h\Z\,$-principal gradation of~$\mathring{\mfg}$ as
\[
\mathring{\mfg}=\bigoplus_{j\in \Z/h\Z}\mr{\mfg}_j,
\]
then we have $e_i\in\mr{\mfg}_{\bar{1}}$,
$f_i\in\mr{\mfg}_{\overline{-1}}$, $h_i\in\mr{\mfg}_{\bar{0}}$
for~$i=0, \dots, n$.

Let $\Ld=\sum_{i=0}^{n}{e}_i$ and~$\mr{\mfs}$ be the centralizer
of~$\Ld$ in~$\mathring{\mfg}$. Then~$\mr{\mfs}$ is a Cartan
subalgebra of~$\mr{\mfg}$. We fix a basis $\{T_j|\,j\in I\}$
of~$\mr{\mfs}$ as
\begin{align*}
T_j=&\Ld^j,\qquad j=1, 3, \ldots, 2n-3, \\
T_{(n-1)'}=&\sqrt{n-1}\,\kappa\Big(e_{n,1}-\frac{1}{2}e_{n+1,1}-\frac{1}{2}e_{n,2n}+\frac{1}{4}e_{n+1,2n}\\
&\quad+(-1)^n\big(e_{2n,n+1}-\frac{1}{2}e_{2n,n}-\frac{1}{2}e_{1,n+1}+\frac{1}{4}e_{1,n}\big)\Big)
\end{align*}
where $\kappa=1$ (resp. $\sqrt{-1}$) when~$n$ is even (resp. odd), and $I$ is the set of exponents of $\mr\mfg$ given by
\[
I=\{1, 3, 5,\ldots,2n-3\}\cup\{(n-1)'\}.
\]
Here $(n-1)'$ indicates that when $n$ is even the multiplicity of
the exponent $n-1$ is $2$. These matrices $T_j$ belong to
$\mathring{\mfg}_j$ respectively, and satisfy
\[
(T_i|T_{h-j})=(n-1)\dt_{i,j}.
\]

To construct the desired bases, we need the root space decomposition
of~$\mr{\mfg}$ with respect to~$\mr{\mfs}$. Note that the set of
eigenvalues of~$\Lambda$ is
\[
\{\om\in\C\mid\om^h=1\}\cup\{0\},
\]
in which the multiplicity of $0$ is $2$.
We choose the eigenvectors $\eta_{\omega}, \eta_0, \eta_{0'}$ associated to eigenvalues $\omega$, $0$ respectively as follows
\begin{align*}\label{}
&\eta_\om=(\frac{1}{2}, \om^{-1}, \ldots, \om^{-(n-1)},\ \frac{1}{2}\om^{n-1}, \om^{n-2},\ldots,\om,1)^t,\\
&\eta_{0}=(-\frac{1}{2}\psi_1+\psi_{2n})+\kappa^{-1}(\psi_{n}-\frac{1}{2}\psi_{n+1}),\\
& \eta_{0'}=(-\frac{1}{2}\psi_1+\psi_{2n})-\kappa^{-1}(\psi_{n}-\frac{1}{2}\psi_{n+1}),
\end{align*}
where $\psi_i$ is the~$2n$-dimensional column vector with the $i$-th
entry being $1$ and all other entries being zero, and $\cdot^t$ is
the usual transposition of matrices. These eigenvectors give a
 common eigenspace decomposition for $T_j\ (j\in I)$:
\begin{align*}\label{}
&T_j\,\eta_\al=\al^j\,\eta_\al,\quad j=1,3,\dots,2n-2, \\
&T_{(n-1)'}\,\eta_\al=\big((-1)^{n-1}\dt_{\al,0}+(-1)^{n}\dt_{\al,0'}\big)\sqrt{n-1}\,\eta_\al.
\end{align*}
Introduce a map~$\sg:\,\C^{2n\times 2n}\to\mathring{\mfg}$,
$A\mapsto A-S A^T S$, and define the $2n\times 2n$ matrices
\begin{equation*}\label{}
A_{(\al,\beta)}=\sg(\eta_\al\eta_{-\beta}^T),
\end{equation*}
where $\al, \beta$ are eigenvalues of~$\Ld$. These matrices satisfy
\begin{align*}\label{}
&[T_j, A_{(\al,\beta)}]=(\al^j+\beta^j)A_{(\al,\beta)},\quad j=1,3,\ldots,2n-3,\\
&[T_{(n-1)'},
A_{(\al,\beta)}]=(\dt_{\al,0}-\dt_{\al,0'}+\dt_{\beta,0}-\dt_{\beta,0'})\sqrt{n-1}\,A_{(\al,\beta)},
\end{align*}
from which one can obtain the root space decomposition of~$\mr{\mfg}$ with respect to~$\mr{\mfs}$.

Now denote by~$A_{(\al,\beta),j}$ the homogeneous components of~$A_{(\al,\beta)}$ in $\mr{\mfg}_j$, and fix~$\om=\exp\big(2\pi i/h\big)$.
One can verify the following relations
\begin{align*}\label{}
(A_{(1,\om^r),0}|A_{(-1,-\om^s),0})&=-h\dt_{r,s},\\
(A_{(1,\om^r),0}|A_{(-1,\al),0})&=0,\\
(A_{(1,\al),0}|A_{(-1,\beta),0})&=2(1-\dt_{\al,\beta}),
\end{align*}
where~$1\leq r,s\leq n-2$ and~$\al,\beta\in\{0,0'\}$. According to
these relations, we choose two bases of $\mr{\mfg}$:
\begin{align*}
&\{T_j\mid j\in I\}\cup\{\tilde{X}^{(r)}_{m}\mid r=1, \dots, n;\ m\in \Z/h\Z\},\\
&\{T_j\mid j\in I\}\cup\{\tilde{Y}^{(r)}_{m}\mid r=1, \dots, n;\
m\in \Z/h\Z\},
\end{align*}
\begin{equation}\label{XrYr}
\begin{array}{cccc}
\hline
& 1\leq r\leq n-2 & r=n-1 & r=n \\
\hline
\tilde{X}^{(r)}_{m} :& \frac{1}{\sqrt{h}}{A}_{(1,\om^r),m} &\frac{1}{\sqrt{2}}{A}_{(1,0),m} & \frac{1}{\sqrt{2}}{A}_{(1,0'),m} \\
\tilde{Y}^{(r)}_{m} :&-\frac{1}{\sqrt{h}}{A}_{(-1,-\om^r),m} & \frac{1}{\sqrt{2}} {A}_{(-1,0'),m} & \frac{1}{\sqrt{2}}{A}_{(-1,0),m}\\
\hline
\end{array}
\end{equation}

The above two bases of~$\mr\mfg$ help us to construct a pair of dual
bases~\eqref{dual}, \eqref{dual2} of the affine Lie algebra~$\mfg$
that satisfy~\eqref{X0Y0}-\eqref{XY}. We use the principal
realization of~$\mfg$~\cite{Kac}
\[\mfg=\bigoplus_{m\in\Z}\ld^m\mr\mfg_{\bar{m}}\oplus\C c\oplus\C d.\]
Note that the set of exponents of~$\mfg$ is~$E=I+h\,\Z$, and the
principal Heisenberg subalgebra is generated by
\[H_{j}=\sqrt{2}\,\ld^j\,T_{\bar{j}},\ j\in E.
\]
The two bases \eqref{dual}, \eqref{dual2} of~$\mfg$ can be chosen as
\begin{align*}
&\frac1{\sqrt{h}}H_j,\ X^{(r)}_m=\lambda^{m}\tilde{X}^{(r)}_{\bar{m}},\ c,\ d;\\
&\frac1{\sqrt{h}}H_{-j},\
Y^{(r)}_{-m}=\lambda^{-m}\tilde{Y}^{(r)}_{\overline{-m}},\ d,\ c
\end{align*}
with the coefficients~$\beta_{r,j}$ that appear in~\eqref{beta}
given by
\begin{align}\label{betarj1}
\beta_{r,j}=&\left\{\begin{array}{ll}
                      \sqrt{2}( 1+\om^{rj}), & r=1, 2, \ldots, n-2,\ j\ne(n-1)',\\
                      \sqrt{2}, & r=n-1, n,\ j\ne(n-1)',\\
                      \sqrt{2n-2}(\dt_{r,n-1}-\dt_{r,n}), & j=(n-1)'.
                   \end{array}\right.
\end{align}

To write down the Kac-Wakimoto bilinear equations~\eqref{kwhr}, we
still need to compute the constants
$g_r=(\mr{\rho}^\vee|\mr{X}^{(r)}_0)(\mr{\rho}^\vee|\mr{Y}^{(r)}_0)$.
Note that in the principal realization of $\mfg$, the Weyl
generators are given by
\[\tilde{e}_i=\ld\,e_i,\ \tilde{f}_i=\ld^{-1}f_i,\ \alpha_i^{\vee}=h_i+\frac{c}{h},~~ i=0,\dots, n,\]
so we have
\[(\mr{\rho}^\vee|\mr{X}^{(r)}_0)=\left(\mr{\rho}^\vee\left|X^{(r)}_0+\frac{c}{h}\sum_{i=1}^n a_i\right.\right)=\sum_{i=1}^n a_i,\]
where $a_i$ are the coefficients in the following linear expansion
\[X^{(r)}_0=\sum_{i=1}^n a_i\,h_i=\sum_{i=1}^n a_i\,\left(\alpha_i^{\vee}-\frac{c}{h}\right) \in \mr{\mfg}_0.\]
According to the realization~\eqref{efhh}-\eqref{efht}, given any
\[\diag(b_1,b_2,\dots,b_{2n})=\sum_{i=1}^n a_i\,h_i\in\mathring{\mfg}_0,\]
the summation $\sum_{i=1}^n a_i$ reads
\[\sum_{i=1}^n a_i=-\sum_{i=1}^{n-1}(n-i)b_i.\]
By using this formula, we obtain
\begin{equation}\label{ggm}
g_r=\left\{\begin{array}{ll}
\frac{n-1}{2}\frac{2-\om^r-\om^{-r}}{2+\om^r+\om^{-r}}, & r=1,\ldots,n-2, \\
\frac{(n-1)^2}{2} & r=n-1, n.
\end{array}\right.
\end{equation}

\begin{prop}
The constants $g_r$ and $\beta_{r,j}$ in the Kac-Wakimoto hierarchy
of bilinear equations \eqref{kwhr} for $D_n^{(1)}$ are given by
\eqref{betarj1} and \eqref{ggm}.
\end{prop}
Note that the values $\beta_{r,j}$ depend on the choice of the dual
bases~\eqref{dual}, \eqref{dual2}. However, it is easy to see that
the constants $g_r$ are independent of the choice of such bases.

\vskip 2ex

In \cite{GM}, Givental and Milanov proved that the total descendant
potential for semisimple Frobenius manifolds associated to a simple
singularity satisfies the following hierarchy of Hirota bilinear
equations:
\begin{equation}\label{gmbl}
\begin{split}
\res_{z=0}&z^{-1}\sum_{r=1}^n g_r e^{\sum_{j\in
E_+}2\beta_{r,\bar{j}}\,z^j y_j} e^{-\sum_{j\in
E_+}\,\beta_{r,\overline{-j}}\,z^{-j}\pd_{y_j}/j}\tau(t+y)\tau(t-y)\\
&=\Big( 2h\sum_{j\in E_+}j\,y_j\pd_{y_j}+\frac{n
h(h+1)}{12}\Big)\tau(t+y)\tau(t-y),
\end{split}
\end{equation}
where the coefficients $\beta_{r,j}$ are the same as in
\eqref{kwhr}, and~$g_r$ are given explicitly in~\cite{GM}. By
comparing the constants $g_r$ \eqref{ggm} with those in~\cite{GM},
we obtain the following corollary.
\begin{cor}
The hierarchy~\eqref{gmbl} for the~$D_n$ singularity coincides with
the Kac-Wakimoto hierarchy of type~$D_n^{(1)}$ associated to the basic representation and its principal vertex operator construction.
\end{cor}
Namely, we conform Givental and Milanov's conjecture~\cite{GM} for the case~$D_n$.

\section{Concluding remarks}

We study in \cite{WZ} the tau structure of the Drinfeld-Sokolov
hierarchy associated to $D_n^{(1)}$ and the zeroth vertex of its
Dynkin diagram following the approach of \cite{DZ}. So we can define
the tau function by using the tau symmetry of the Hamiltonian
structures, and establish the equivalence between this definition of
the tau function for this hierarchy and that given by Hollowood and
Miramontes~\cite{HM}. Basing on the tau structure, we plan to show
that this Drinfeld-Sokolov hierarchy coincides with the
bihamiltonian integrable hierarchy constructed according to the
axiomatic scheme developed by Dubrovin and Zhang \cite{DZ} on the
formal loop space of the semisimple Frobenius manifold associated to
the $D_n$-type Weyl group. This assertion together with the result
of this note would imply that Givental's total descendant potential
associated to the $D_n$ singularity is a tau function of Dubrovin
and Zhang's hierarchy.

While we prepared to do an analogous computation for the cases
$E_7$, $E_8$ of Givental and Milanov's conjecture~\cite{GM}, we
learned from \cite{FJR} that Frenkel, Givental and Milanov have
obtained a proof of this conjecture in general. We hope however that
this short note might be helpful to a better understanding of the
relationship between Givental's total descendant potentials and
integrable systems.

\vskip 0.4truecm \noindent{\bf Acknowledgments.} The author would like to thank
Boris Dubrovin, Si-Qi Liu and Youjin Zhang for advises, he would also like to
thank Todor Milanov for helpful comments.
This work is partially supported by the National Basic Research Program of China (973 Program)  No.2007CB814800.

\end{document}